\documentclass[a4paper,11pt]{article}
\pdfoutput=1 

\usepackage{jinstpub} 

\title{\boldmath Eikonal Approximation and the Conical Effect in Smith-Purcell Radiation}

\author[a,1]{V.V. Syshchenko,\note{Corresponding author.}}
\author[a]{A.I. Tarnovsky}

\affiliation[a]{Belgorod State University,\\Pobedy Street, 85, Belgorod 308015, Russian Federation}

\emailAdd{syshch@yandex.ru}

\abstract{The diffraction and transition radiation under normal incidence of a particle to the semi-infinite dielectric plate had been considered previously using the eikonal approximation in the transition radiation theory. This approach is valid for the high radiation frequencies domain (where the dielectric permittivity of the plate material is close to unit), including X-ray domain. In the present report this method is applied to the case of oblique incidence of the particle on the plate (but parallel to the plate's edge) as well as to the periodic set of such plates (Smith-Purcell radiation). The oblique incidence of the particle (both in the case of the single plate and in the case of the periodic grating of such plates) leads to the so called conical effect in the radiation angular distribution observed recently. We outline that the origin of this effect lies in the superluminal motion along the plate's edge of the disturbance produced by the incident particle's field in the plates. The analogous effects arise in various physical situations. Also we compare our results to ones obtained using different technique.}


\arxivnumber{} 

\proceeding{XIII$^{\text{th}}$ International Symposium <<Radiation from Relativistic Electrons in Periodic Structures>>\\
  September 16-20, 2019\\
  Belgorod, Russian Federation}

\begin{document}
\maketitle
\flushbottom

\section{Introduction}
\label{sec:intro}

Transition radiation permits precise description only in the case of the incident particle's interaction with the simplest target that is the uniform infinite plate. All other cases need some approximate methods. Some of that methods are briefly described in the sections \ref{Born.method}--\ref{eikonal.method}.

The diffraction and transition radiation (DR and TR) under normal incidence of a particle to the semi-infinite dielectric plate had been considered in \cite{jof.2010} using the eikonal approximation in the transition radiation theory \cite{arsa.2006, pov.2007}. This approach is valid for the high radiation frequencies domain (where the dielectric permittivity of the plate material is close to unit), including X-ray domain. In the present paper this method is applied to the case of oblique incidence of the particle on the plate (but parallel to the plate's edge) as well as to the periodic set of such plates (Smith--Purcell radiation). The oblique incidence of the particle (both in the case of the single plate and in the case of the periodic grating of such plates) leads to the so called conical effect in the radiation angular distribution recently observed in \cite{Naumenko}. We outline that the origin of this effect lies in the superluminal motion along the plate's edge of the disturbance produced by the incident particle's field in the plates. The analogous effects arise in various physical situations \cite{Bolot.Ginz, Davydov, Syshch12} (section \ref{interpretation}). 

The TR spectral-anguylar density on such plate is calculated in the section \ref{TR.section}. Also we compare our results to ones of \cite{Tishch.2004, Tishch.2} obtained using different technique (section \ref{discussion}).

\section{Eikonal and other approximations in TR and DR theory}
\label{sec:method}

The spectral-angular density of the diffraction or transition radiation 
can be written in the form \cite{arsa.2006, pov.2007}
\begin{equation}
    \label{spectral.angular.0}
    \frac{d\mathcal E}{d\omega d\Omega} = \frac{\omega^2}{(8\pi^2)^2 c}
    \left|\mathbf k\times\mathbf I\right|^2,
\end{equation}
where
\begin{equation}
    \label{I.determination}
    \mathbf I= \int \left(1-\varepsilon_\omega(\mathbf r) \right)
    \mathbf E_\omega(\mathbf r) \, e^{-i\mathbf k\mathbf r} \, d^3r \,,
\end{equation}
$\mathbf k$ is the wave vector of the radiated wave, $|\mathbf k| = \omega /c$,
\begin{equation}
    \mathbf E_\omega (\mathbf r) = \int_{-\infty}^\infty \mathbf E
    (\mathbf r,t)e^{i\omega t} dt
\end{equation}
is the Fourier component by time of the electric field produced by the moving particle in the
substance of the target with the dielectric permittivity
$\varepsilon_\omega(\mathbf r)$.

\subsection{Born approximation}\label{Born.method}

If $\left| 1-\varepsilon_\omega(\mathbf r) \right| \ll 1$, the
precise value of the field in the target in (\ref{I.determination}) could be replaced
in the first approximation by non-disturbed Coulomb field of the
uniformly moving particle in vacuum
\begin{equation}
    \label{Coulomb.Fourier}
    \mathbf E^{(Coulomb)}_\omega (\boldsymbol\rho , z) = \frac{2e\omega}{v^2 \gamma}\, e^{i(\omega/ v)z}
    \left\{ \frac{\boldsymbol\rho}{\rho} K_1  \left(\frac{\omega \rho}{v \gamma} \right) -
    i \frac{\mathbf v}{v \gamma} K_0  \left(\frac{\omega \rho}{v \gamma} \right) \right\} \,,
\end{equation}
where $K_n(x)$ is the modified Bessel function of the second kind (McDonald function), $\boldsymbol\rho$ is the component of $\mathbf r$ perpendicular to $\mathbf v$, $z$ is the component parallal to $\mathbf v$, $\gamma = \sqrt{1-v^2/c^2}$ is the particle's Lorentz factor. The resulting approximation is analogous to Born approximation in the quantum theory of scattering.

Although this approximation permits to calculate the TR characteristics for the targets with complex geometry \cite{Syshch12}, its applicability is restricted
by the range of extremely high frequencies,
\begin{equation}
    \label{Born.condition}
    \omega \gg \gamma\omega_p \,.
\end{equation}
For investigation of radiation in the range of more soft photons, some
different approximate method is needed.

\subsection{Durand approximation}\label{Durand.method}

The simple modification of the above approach was proposed by L. Durand in \cite{Durand}; it consists in the substitution $\mathbf k \to \sqrt{\varepsilon_\omega}\, \mathbf k$ in (\ref{spectral.angular.0}) and (\ref{I.determination}) 
\begin{equation}
    \label{spectral.angular.D}
    \frac{d\mathcal E}{d\omega d\Omega} = \frac{\omega^2}{(8\pi^2)^2 c}
    \left|\sqrt{\varepsilon_\omega}\, \mathbf k\times\mathbf I^{(Durand)} \right|^2,
\end{equation}
\begin{equation}
    \label{I.D}
    \mathbf I^{(Durand)}= \int \left(1-\varepsilon_\omega(\mathbf r) \right)
    \mathbf E^{(Coulomb)}_\omega(\mathbf r) \, e^{-i\sqrt{\varepsilon_\omega}\, \mathbf k\mathbf r} \, d^3r \,.
\end{equation}

\subsection{Eikonal approximation}\label{eikonal.method}

A simple variant of eikonal approximation in TR theory was
developed in \cite{arsa.2006, pov.2007}. In that approach the Coulomb field of the incident particle (\ref{Coulomb.Fourier}) is approximated by the packet of the transverse waves 
\begin{equation}
    \label{paket}
    \mathbf E_\omega (\mathbf r)  = \frac{2e\omega}{v^2\gamma} \, e^{i\omega z/c} \,
    \frac{\boldsymbol\rho}{\rho} \,
    K_1\left( \frac{\omega\rho}{v\gamma}\right) \,,
\end{equation}
and account of the gradual evolution of this field inside the target leads to the following formula for the $\mathbf I$ (\ref{I.determination}) component that perpendicular to the particle's velocity $\mathbf v$:
\begin{equation}
    \label{I.eik}
    \mathbf I^{(eik)}_\perp = \frac{2e\omega}{v^2\gamma} \int d^3 r\, e^{i(\omega/c -k_z)z} \,
    e^{-i\mathbf k_\perp\boldsymbol\rho}\, \frac{\boldsymbol\rho}{\rho}
    K_1\left( \frac{\omega\rho}{v\gamma}\right)
    \exp\left[-i \frac{\omega}{2c} \int_{-\infty}^z
    (1-\varepsilon_\omega(\mathbf r))\,dz\right] ,
\end{equation}
The eikonal approximation in DR and TR theory is valid for
\begin{equation}
    \omega\gg\omega_p
\end{equation}
(compare with the condition (\ref{Born.condition})). For small angles between $\mathbf k$ and $\mathbf v$,
\begin{equation}\label{angle.restriction.0}
    \theta\ll \sqrt{c/L\omega},
\end{equation}
where $L$ is the thickness of the target along the particle's velocity $\mathbf v$, the first exponent in (\ref{I.eik}) can be replaced by the unit, and the integration over $z$ in (\ref{I.eik}) could be performed in general form (that had been done in \cite{jof.2010}). However, in the problem under consideration the integration over  $z$ in (\ref{I.eik}) can be performed without the restriction (\ref{angle.restriction.0}). 

Eq. (\ref{I.eik}) had been used in \cite{jof.2010} for description of the radiation arising under normal incidence of the particle on the semi-infinite uniform plate (as well as the set of parallel plates). Here we consider the case of oblique incidence.

\begin{figure}[htbp]
\vspace*{-2mm}\centering
\includegraphics[width=0.8\textwidth]{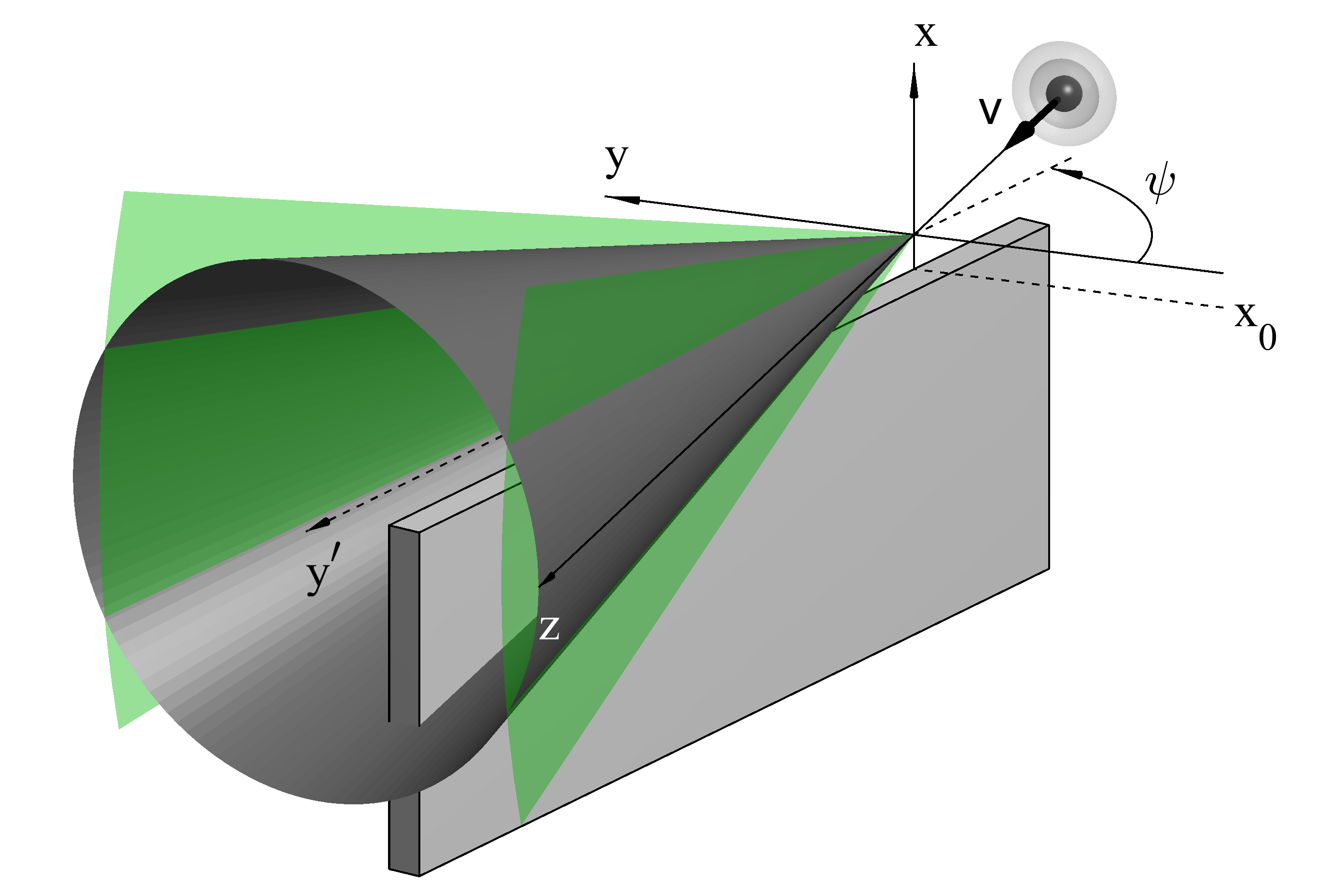} 
\caption{\label{geometry.1} Geometry of the problem. The arising radiation is concentrated near the gray cone with the semi-apex angle $\pi/2 - \psi$ and the axis along the plate's rib. The parts of the cones that (bound border restrict) the domains (\ref{poln.vnut}) are pictured in green.}
\end{figure}

Let the particle is incident on the semi-infinite dielectric plate with the thickness $a$ and uniform dielectric permittivity $\varepsilon_\omega$. The particle's velocity is parallel to the edge of the plate and makes the angle $\psi$ with the normal to the plate's surface (figure \ref{geometry.1}). It is convenient to choose the $z$ axis along the particle's velocity $\mathbf v$, and then the particle's flight above the plate's edge corresponds to negative values of the impact parameter $x_0$ while the flight through the plate to positive ones.

Inegration over $z$ in (\ref{I.eik}) gives
\begin{equation}
    \label{I.eik.plate.0}
    \mathbf I^{(eik)}_\perp = \frac{2e\omega}{v^2\gamma} (1-\varepsilon_\omega)
    \frac{\exp\left\{ i\frac{a}{\cos\psi} \left( \frac{\omega}{c} - k_z - \frac{\omega}{2c} (1-\varepsilon_\omega) \right) \right\} - 1}{i \left( \frac{\omega}{c} - k_z - \frac{\omega}{2c} (1-\varepsilon_\omega) \right)} 
\end{equation}
\begin{equation*}
    \times \int \Theta(x_0-x) \,
    \underbrace{ \frac{x\mathbf e_x + y \mathbf e_y}{\sqrt{x^2 + y^2}}
    K_1\left( \frac{\omega}{v\gamma} \sqrt{x^2 + y^2} \right) } \, 
    \exp\left\{i\left[ \left( \frac{\omega}{c} -k_z \right) \tan\psi - k_y \right] y\right\} \, \exp ( -ik_x x) \, dxdy \,,
\end{equation*}
where $\Theta (\xi)$ is the Heaviside's step function ($\Theta (\xi) = 0$ for $\xi < 0$ and  $\Theta (\xi) = 1$ for $\xi\geq 0$).

\subsection{Interpretation of the conical effect}\label{interpretation}

Already at this stage we can see and discuss some features of the arising radiation. The factor marked by the underbrace disappears from (\ref{I.eik.plate.0}) in the case of incidence of the plane wave $\mathbf E_\omega (\mathbf r) = \mathbf E_0 e^{i\,\mathbf k^{(i)} \,\mathbf r} = \mathbf E_0 e^{i \omega z/c}$ instead of the wave packet (\ref{paket}), and the integration over $y$ would give the $\delta$-function
\begin{equation}\label{delta.function.0}
\delta \left[ \left( \frac{\omega}{c} -k_z \right) \tan\psi - k_y \right] \,.
\end{equation}
The argument of this $\delta$-function becomes zero when
\begin{equation}\label{delta.function.1}
k_z \sin\psi + k_y \cos\psi = \frac{\omega}{c} \sin\psi \,.
\end{equation}
The value on the left is the projection of the radiated wave vector $\mathbf k$ on the axis $y'$ parallel to the plate's edge (figure \ref{geometry.1}):
\begin{equation}\label{delta.function.2}
k_z \sin\psi + k_y \cos\psi = k_{y'} \,,
\end{equation}
so the equation (\ref{delta.function.1}) means
\begin{equation}\label{delta.function.3}
k_{y'} = \frac{\omega}{c} \sin\psi \,.
\end{equation}
But the value $(\omega/c) \sin\psi$ can be interpreted as the projection of the incient wave vector  $\mathbf k^{(i)}$ (parallel to  $\mathbf v$) on the same axis $y'$:
\begin{equation}\label{delta.function.4}
k_{y'} =  k^{(i)}_{y'} \,.
\end{equation}
The last equality together with the equality of the lengths of the wave vectors,  $|\mathbf k| = \omega/c = |\mathbf k^{(i)}|$, leads to the conclusion that the arising radiation would be distributed over the conical surface with the main axis $y'$ and the half-opening angle $\psi$ (see figure  \ref{geometry.1}). 

\begin{figure}[htbp]
\begin{center}
\includegraphics[width=0.6\textwidth]{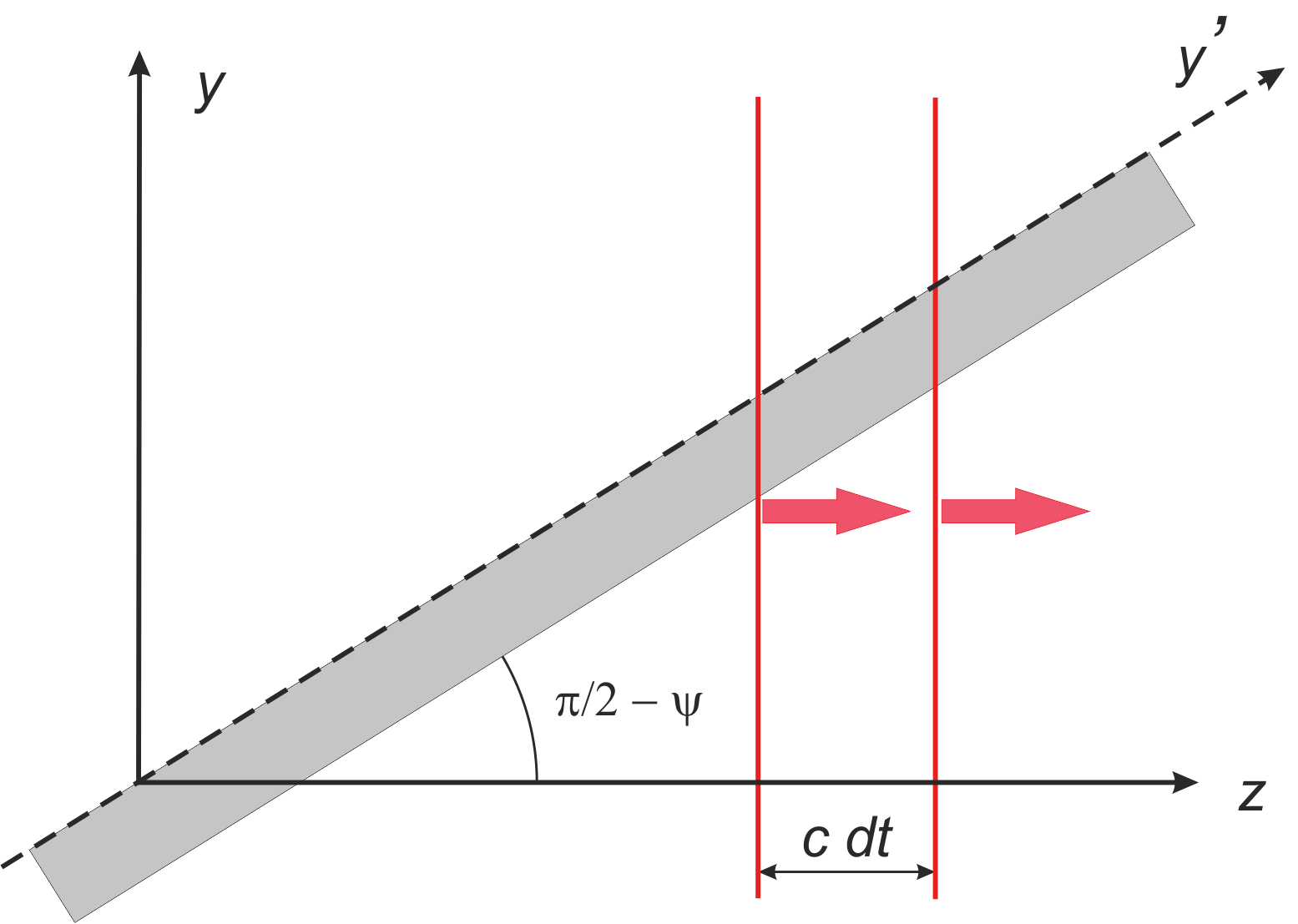}
\end{center}
\vspace*{-5mm}\caption{\small The intersection of the wave front (red vertical line) travelling along the $z$ axis with the velocity $c$ with the target will move along the last one with the velocity $c/\sin\psi > c$.}\label{superluminal}
\end{figure}

This conical character of the angular distribution of the radiation under oblique incidence on the plate permits clear interpretation as the manifestation of Cherenkov mechanism. Consider the plane wave incident on our plate under the angle $\pi/2-\psi$ (figure \ref{superluminal}). This wave creates in the plate the perturbation that travels along the plate's edge with the superluminal speed
\begin{equation}
    \label{25}
    v_{\mbox{\small perturbation}} = \frac{\omega}{k^{(i)}_\parallel} = \frac{c}{\cos(\pi/2-\psi)} = \frac{c}{\sin\psi} > c \,.
\end{equation}
Indeed, while the incident wavefront travels the distance $c\,dt$ in the $z$ direction, the point of intersection of the wave front and the plate's edge travels the distance
\begin{equation}
    \label{26}
    \frac{z + c\,dt}{\sin\psi} - \frac{z}{\sin\psi} = \frac{c\,dt}{\sin\psi}
\end{equation}
along the $y'$ axis, so the perturbation moves along the edge with the velociy  $c/\sin\psi > c$. This superluminal motion leads to generation of the radiation analogous to Cherencov one.  The half-opening angle of the Cherenkov cone determined by the relation $\cos\theta_{\mbox{\footnotesize Ch}} = c/v_{\mbox{\small perturbation}}$ is just equal to $\pi/2-\psi$:
\begin{equation}
    \label{27}
    \cos\theta_{\mbox{\footnotesize Ch}} =  \frac{c}{v_{\mbox{\small perturbation}}} =\cos(\pi/2-\psi) \,.
\end{equation}
This interpretation clearly demonstrates why the cone opening angle depends neither on the radiation frequency nor on the dielectric properties of the plate's material.

The analogous cone effects had been discussed in a large variety of physical situations (see, e.g., \cite{Bolot.Ginz, Davydov, Syshch12}). 

The presence of the underbraced factor in (\ref{I.eik.plate.0}) will lead to the blur of the cone, see below.

\section{Diffraction radiation}

\begin{figure}[htbp]
\vspace*{-2mm}\centering
\includegraphics[width=\textwidth]{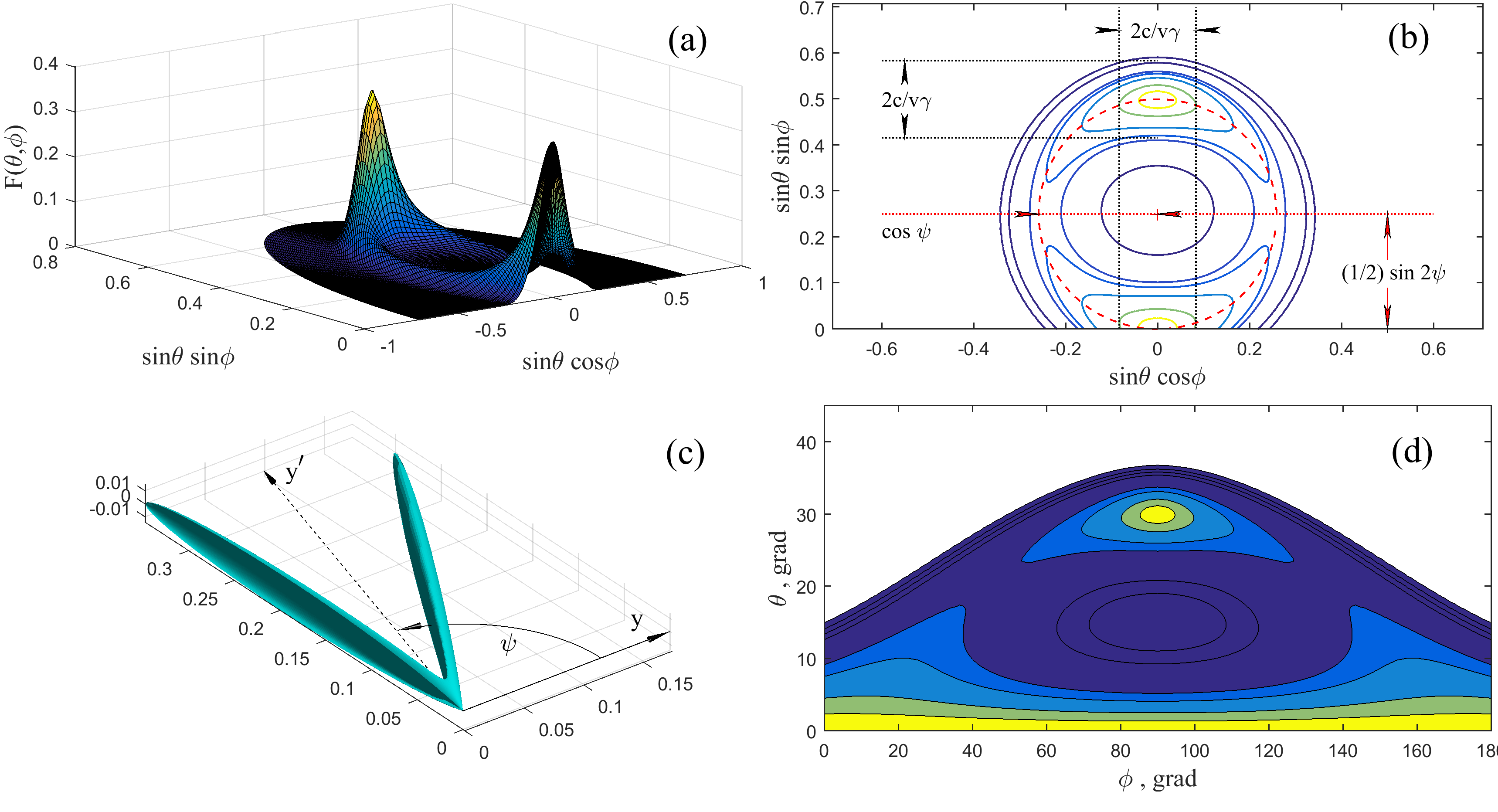}
\vspace*{-2mm}\caption{\label{angular.1} The value $F(\theta ,\varphi)$ that decribes the anguar distribution of DR acccording to (\ref{spectral.angular.2}) under $\gamma = 12$, $x_0 \omega / c\gamma = -3$, $\psi = 75^\circ$, presented as a surface plot (a), as an isolevel contour plot ((b) and (d)), and as a directional diagram (c).}
\end{figure}

\begin{figure}[htbp]
\vspace*{-2mm}\centering
\includegraphics[width=\textwidth]{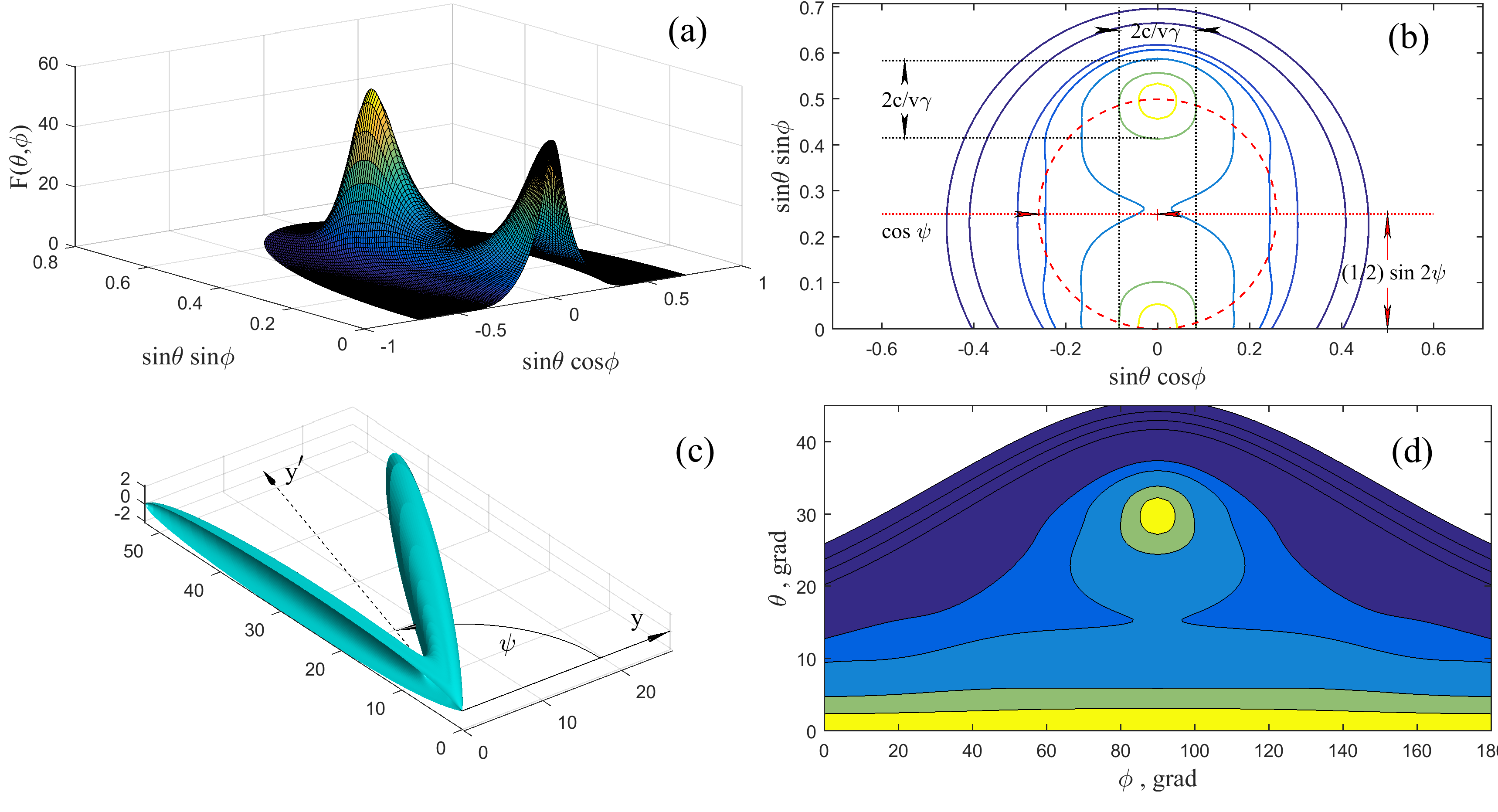}
\vspace*{-2mm}\caption{\label{angular.2} The same as in figure \ref{angular.1} for the case $x_0 \omega / c \gamma = -0.5$.}
\end{figure}

Integrating (\ref{I.eik.plate.0}) over $y$ we obtain
\begin{equation}\label{I.eik.plate.2}
    \mathbf I^{(eik)}_\perp = \frac{2\pi e}{v} (1-\varepsilon_\omega)
    \frac{\exp\left\{ i\frac{a}{\cos\psi} \left( \frac{\omega}{c} - k_z - \frac{\omega}{2c} (1-\varepsilon_\omega) \right) \right\} - 1}{i \left( \frac{\omega}{c} - k_z - \frac{\omega}{2c} (1-\varepsilon_\omega) \right)} \, e^{ -i k_x x_0} 
\end{equation}
\begin{equation*}
    \times \,\,
    \frac{ \exp \left\{x_0 \sqrt{  \left(\frac{\omega}{v\gamma}\right)^2 + \left[ \left( \frac{\omega}{c} -k_z \right) \tan\psi - k_y \right]^2} \right\} }
    {-ik_x + \sqrt{  \left(\frac{\omega}{v\gamma}\right)^2 + \left[ \left( \frac{\omega}{c} -k_z \right)\tan\psi - k_y \right]^2}} 
    \left\{ -\mathbf e_x +
    i \frac{ \left( \frac{\omega}{c} -k_z \right) \tan\psi - k_y }
    { \sqrt{  \left(\frac{\omega}{v\gamma}\right)^2 + \left[ \left( \frac{\omega}{c} -k_z \right) \tan\psi - k_y \right]^2}} \,
    \mathbf e_y \right\} .
\end{equation*}
The substitution of (\ref{I.eik.plate.2}) into (\ref{spectral.angular.0}) in the case $x_0 < 0$ leads to the spectral-angular density of DR:
\begin{equation}
    \label{spectral.angular.1}
    \frac{d\mathcal E}{d\omega d\Omega} = \frac{e^2}{8\pi^2 v^2} \, (1-\varepsilon_\omega)^2 \,
    \frac{1-\cos\left[ \frac{a\omega}{c\cos\psi} \left( \frac{1+\varepsilon_\omega}{2} - \varkappa_z \right) \right]}{\left( \frac{1+\varepsilon_\omega}{2} - \varkappa_z \right)^2}
    F(\theta , \varphi) \,,
\end{equation}
where $\boldsymbol{\varkappa} = \mathbf k / |\mathbf k|$ is the unit vector in the $\mathbf k$ direction, $\varkappa_x = \sin\theta\cos\varphi$, $\varkappa_y = \sin\theta\sin\varphi$, $\varkappa_z = \cos\theta$, 
\begin{equation}
    \label{spectral.angular.2}
    F(\theta , \varphi) = \frac{ \exp \left\{2x_0 \frac{\omega}{c}
    \sqrt{  \frac{c^2}{v^2\gamma^2} + [(1 -\varkappa_z)\tan\psi - \varkappa_y]^2} \right\} }
    {\varkappa_x^2 + \frac{c^2}{v^2\gamma^2} + [(1 -\varkappa_z)\tan\psi - \varkappa_y]^2}
    \,
    \frac{ \frac{1-\varkappa_x^2}{v^2\gamma^2/c^2} + (2-\varkappa_\perp^2) [(1 -\varkappa_z)\tan\psi - \varkappa_y]^2}
    {\frac{c^2}{v^2\gamma^2} + [(1 -\varkappa_z)\tan\psi - \varkappa_y]^2} \,,
\end{equation}
$\varkappa_\perp^2 = \varkappa_x^2 + \varkappa_y^2$.
The value $\gamma^{-2} F(\theta, \varphi)$ that describe the angular distribution of radiation is presented in figures \ref{angular.1} and \ref{angular.2}.

The formulae (\ref{I.eik.plate.2}) and (\ref{spectral.angular.2}) demonstrate that the radiation is concentrated near the cone that is determined by the condition (\ref{delta.function.0})--(\ref{delta.function.1}): the argument of the $\delta$-function (\ref{delta.function.0}) is present in (\ref{I.eik.plate.2}) and (\ref{spectral.angular.2}) in the square brackets. The main contribution to the radiation is made by the range of values 
\begin{equation*}
\left| (1 -\varkappa_z)\tan\psi - \varkappa_y \right| \lesssim \frac{c}{v\gamma} \,,
\end{equation*}
so the characteristic ``thickness'' of the cone is $\sim 2c/v\gamma$. The exponential factor demonstrates that the increase of $|x_0|$ leads to the decrease of the total intensity, but, however, to the increase of the cone's sharpness  (compare figures \ref{angular.1} and \ref{angular.2}). 

On the other hand, the denominator in (\ref{I.eik.plate.2}) and (\ref{spectral.angular.2}) leads to supression of the radiation for large $|\varkappa_x|$, with the same effective values
\begin{equation*}
\left| \varkappa_x \right| \lesssim \frac{c}{v\gamma} \,,
\end{equation*}
that leads to the persence of two ``horns'' on the angular distribution.

\subsection{DR from the grating}

Consider now the radiation arising under the incidence on the stack of the parallel plates with the period $b$. The contribution into $\mathbf I^{(eik)}_\perp$ from the every next plate will distinct from the contribution from the previous one by the phase factor
\begin{equation*}
    \exp \left[ i\,\omega \frac{b}{\cos\psi} \left( \frac{1}{v} - \frac{\cos\theta}{c} \right) \right] .
\end{equation*}
The radiation in this case will be described (neglecting the emitted radiation refraction in the following plates) by the formula (\ref{spectral.angular.1}) with the additional factor
\begin{equation}
    \label{Smith.Purcell.1}
    2\pi N \sum_{j=-\infty}^\infty \delta \left\{ \omega \frac{b}{\cos\psi} \left( \frac{1}{v} - \frac{\cos\theta}{c} \right) -2\pi j \right\},
\end{equation}
where $N$ is the total number of the plates, $N\gg 1$. These $\delta$-functions mean that the radiation under the angle $\theta$ takes the place only for the frequencies that satisfy the condition
\begin{equation}
    \label{Smith.Purcell.2}
    \omega_j = \frac{v}{\displaystyle 1 - \frac{v}{c} \cos\theta} \cos\psi \frac{2\pi j}{b} \end{equation}
($j$ is the positive integer). This is well known Smith--Purcell condition \cite{Smith.Purcell}.

\section{Transition radiation}\label{TR.section}

When the particle's trajectory crosses the plate ($x_0 > 0$) the integration over $x$ in (\ref{spectral.angular.0}) gives
\begin{equation*}
    \mathbf I^{(eik)}_\perp = \frac{2\pi e}{v} (1-\varepsilon_\omega)
    \frac{\exp\left\{ i\frac{a}{\cos\psi} \left( \frac{\omega}{c} - k_z - \frac{\omega}{2c} (1-\varepsilon_\omega) \right) \right\} - 1}{i \left( \frac{\omega}{c} - k_z - \frac{\omega}{2c} (1-\varepsilon_\omega) \right)} \,
    \frac{1}
    {k_x^2 +  \left(\frac{\omega}{v\gamma}\right)^2 + [( \frac{\omega}{c} -k_z)\tan\psi - k_y]^2} 
\end{equation*}
\begin{equation*}
     \times \left\{ \mathbf e_x \left[ -2ik_x  - \left( -ik_x + \sqrt{  \left(\frac{\omega}{v\gamma}\right)^2 + [( \frac{\omega}{c} -k_z)\tan\psi - k_y]^2} \right) e^{-ik_xx_0} \, e^{-x_0 \sqrt{  \left(\frac{\omega}{v\gamma}\right)^2 + [( \frac{\omega}{c} -k_z)\tan\psi - k_y]^2} }
    \right] \right. 
\end{equation*}
\begin{equation}\label{I.eik.plate.TR}
    \left. \phantom{a} + \mathbf e_y  \left[ 2i \left[ \left( \frac{\omega}{c} -k_z \right) \tan\psi - k_y \right]  \vphantom{\sqrt{\left(\frac{v}{v}\right)^2}}\right.\right.
\end{equation}
\begin{equation*}
\phantom{a}    -i \frac{ ( \frac{\omega}{c} -k_z)\tan\psi - k_y }
    { \sqrt{  \left(\frac{\omega}{v\gamma}\right)^2 + [( \frac{\omega}{c} -k_z)\tan\psi - k_y]^2}} \left( -ik_x + \sqrt{  \left(\frac{\omega}{v\gamma}\right)^2 + [( \frac{\omega}{c} -k_z)\tan\psi - k_y]^2} \right) 
\end{equation*}
\begin{equation*}
    \left. \phantom{a} \times e^{-ik_xx_0} \, e^{-x_0 \sqrt{  \left(\frac{\omega}{v\gamma}\right)^2 + [( \frac{\omega}{c} -k_z)\tan\psi - k_y]^2} }
    \right\} \,.
\end{equation*}
The spectral-angular density of TR is described by the formula (\ref{spectral.angular.1}), however the value $F(\theta , \varphi)$ now is equal to
\begin{equation}
    \label{spectral.angular.3}
    F(\theta , \varphi) = \frac{1}
    {\left( \varkappa_x^2 + \frac{c^2}{v^2\gamma^2} + [(1 -\varkappa_z)\tan\psi - \varkappa_y]^2 \right)^2} 
\end{equation}
\begin{equation*}
    \times \left\{ 4 \left( \varkappa_x^2 + [(1 -\varkappa_z)\tan\psi - \varkappa_y]^2 - \left(-\varkappa_x^2 + \varkappa_y [(1 -\varkappa_z)\tan\psi - \varkappa_y] \right)^2 \right)
    \vphantom{\sqrt{\left(\frac{v}{v}\right)^2}}
    \right. 
\end{equation*}
\begin{equation*}
    +\exp \left\{-2x_0 \frac{\omega}{c}
    \sqrt{  \frac{c^2}{v^2\gamma^2} + [(1 -\varkappa_z)\tan\psi - \varkappa_y]^2} \right\}
\end{equation*}
\begin{equation*}
    \times \left( \varkappa_x^2 + \frac{c^2}{v^2\gamma^2} + [(1 -\varkappa_z)\tan\psi - \varkappa_y]^2 \right)
    \left( 1 - \varkappa_x^2 +(1-\varkappa_y^2)   \frac{[(1 -\varkappa_z)\tan\psi - \varkappa_y]^2}{\frac{c^2}{v^2\gamma^2} + [(1 -\varkappa_z)\tan\psi - \varkappa_y]^2} \right) 
\end{equation*}
\begin{equation*}
    \phantom{a} + 4 \exp \left\{-x_0 \frac{\omega}{c}
    \sqrt{  \frac{c^2}{v^2\gamma^2} + [(1 -\varkappa_z)\tan\psi - \varkappa_y]^2} \right\} 
\end{equation*}
\begin{equation*}
    \times \left( - \varkappa_x^2 (1-\varkappa_x^2) \cos \left(\frac{\omega}{c}\varkappa_x x_0 \right) - \varkappa_x \sqrt{  \frac{c^2}{v^2\gamma^2} + [(1 -\varkappa_z)\tan\psi - \varkappa_y]^2} \, (1-\varkappa_x^2)  \sin\left(\frac{\omega}{c}\varkappa_x x_0 \right)
    \right.
\end{equation*}
\begin{equation*}
    \phantom{a} - (1-\varkappa_y^2) \, [(1 -\varkappa_z)\tan\psi - \varkappa_y]^2  \cos\left(\frac{\omega}{c}\varkappa_x x_0 \right) 
\end{equation*}
\begin{equation*}    
    \phantom{a} -
    \varkappa_x (1-\varkappa_y^2)
    \frac{[(1 -\varkappa_z)\tan\psi - \varkappa_y]^2}{\sqrt{\frac{c^2}{v^2\gamma^2} + [(1 -\varkappa_z)\tan\psi - \varkappa_y]^2}}
    \sin\left(\frac{\omega}{c}\varkappa_x x_0 \right) 
\end{equation*}
\begin{equation*}
    - 2\varkappa_x^2\varkappa_y [(1 -\varkappa_z)\tan\psi - \varkappa_y]^2   \cos\left(\frac{\omega}{c}\varkappa_x x_0 \right) +
    \varkappa_x^3 \varkappa_y
    \frac{[(1 -\varkappa_z)\tan\psi - \varkappa_y]^2}{\sqrt{\frac{c^2}{v^2\gamma^2} + [(1 -\varkappa_z)\tan\psi - \varkappa_y]^2}}
    \sin\left(\frac{\omega}{c}\varkappa_x x_0 \right) 
\end{equation*}
\begin{equation*}
    \left.\left. \phantom{a} - \varkappa_x \varkappa_y \, [(1 -\varkappa_z)\tan\psi - \varkappa_y]^2 \, \sqrt{  \frac{c^2}{v^2\gamma^2} + [(1 -\varkappa_z)\tan\psi - \varkappa_y]^2} \,  \sin\left(\frac{\omega}{c}\varkappa_x x_0 \right)
    \right) \right\} \,.
\end{equation*}
Under $x_0\to\infty$ the second and third groups of terms (that contain exponents) disappear. Hence the first group of terms can be interpreted as describing the transition radiation on the infinite slanted plate. Then the second goup of terms describes the influence from the plate's edge, and the third one describes the interference of these two contributions. 

\begin{figure}[htbp]
\vspace*{-2mm}\centering
\includegraphics[width=\textwidth]{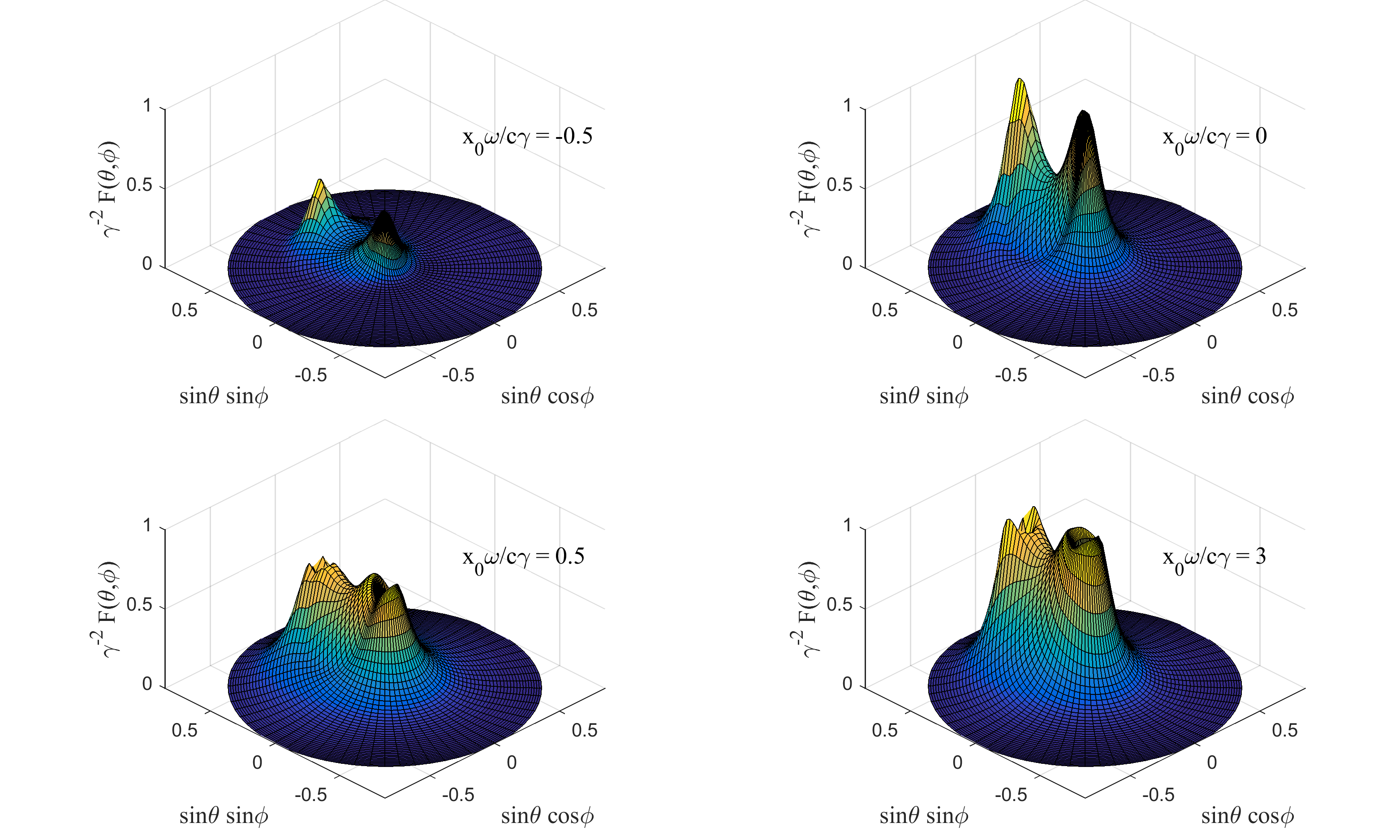}
\vspace*{-2mm}\caption{\label{DR.and.TR} The value $\gamma^{-2} F(\theta ,\varphi)$ that describes the angular distribution of radiation according to (\ref{spectral.angular.2}) for $x_0 \omega / c\gamma = -0.5$ and according to (\ref{spectral.angular.3}) for $x_0 \omega / c\gamma = 0$, 0.5, 3. In all cases $\gamma = 12$, $\psi = 75^\circ$.}
\end{figure}

Plots of the function $\gamma^{-2} F(\theta , \varphi)$ for different values $x_0$ are presented in figure \ref{DR.and.TR}.

\section{Discussion}\label{discussion}

\begin{figure}[htbp]
\vspace*{-2mm}\centering
\includegraphics[width=\textwidth]{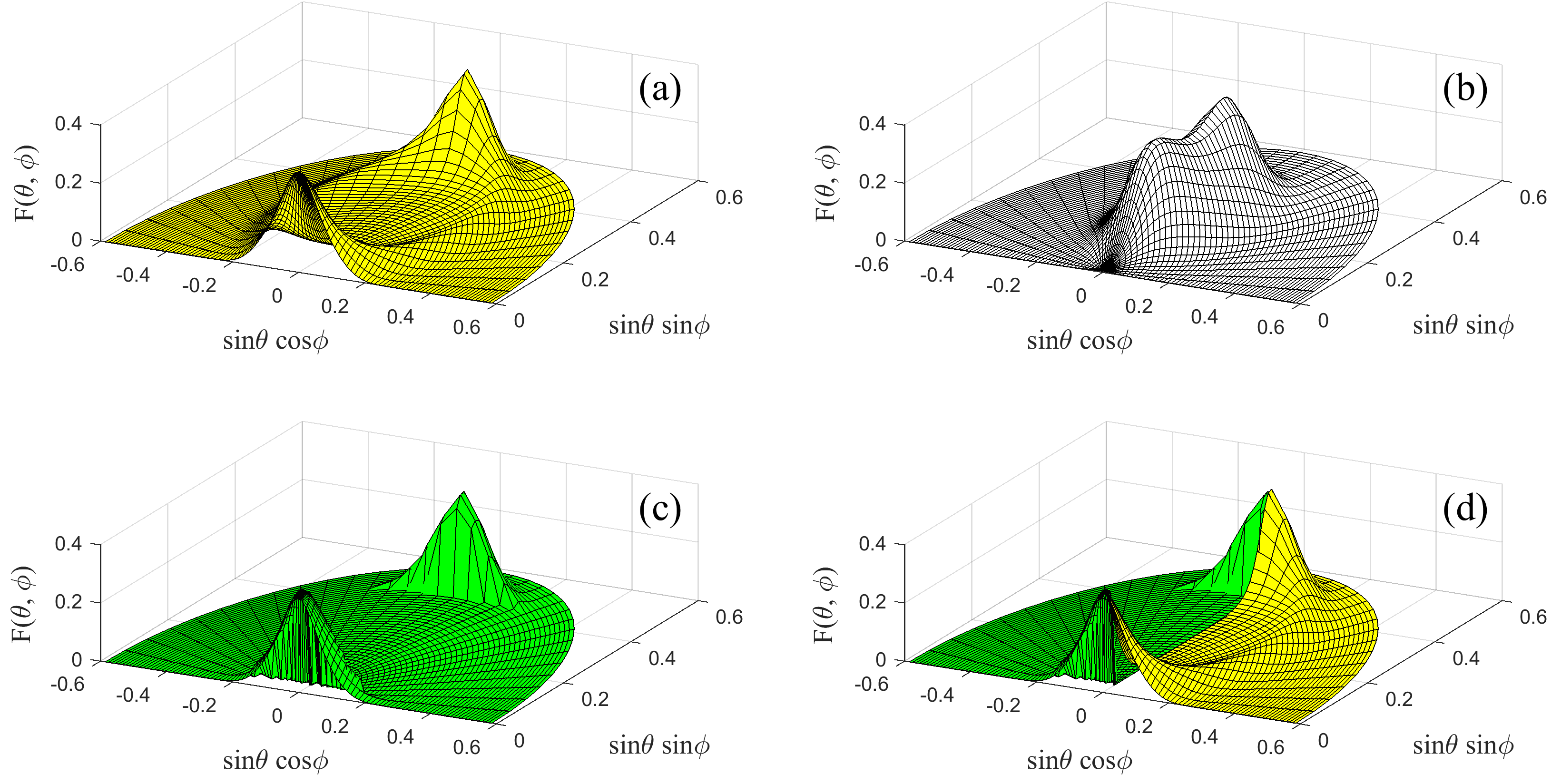}
\vspace*{-2mm}\caption{\label{our.vs.Tishch} The value $F(\theta ,\varphi)$ that describes the angular distribution of DR for $x_0 \omega / c\gamma = -3$,  $\gamma = 12$, $\psi = 75^\circ$,  $\varepsilon_\omega = 0.95$ according to  (\ref{I.eik.plate.0}), (\ref{spectral.angular.2}), that is the same as in figure \ref{angular.1} (a), (\ref{I.plate.Durand}) (b),  (30) from \cite{Tishch.2} (c). The results (a) and (c) are compared in the panel (d).}
\end{figure}

Compare our result for DR under oblique incidence on the semi-infinite plate (\ref{I.eik.plate.2})--(\ref{spectral.angular.2}) with the result obtained in \cite{Tishch.2004, Tishch.2} for the same problem using Durand approach (see section \ref{Durand.method}). Equation (\ref{I.D}) gives us
\begin{equation*}
    \mathbf I^{(Durand)}_\perp = \frac{2\pi e}{v} (1-\varepsilon_\omega)
    \frac{\exp\left\{ i\frac{a}{\cos\psi} \left( \frac{\omega}{v} - \sqrt{\varepsilon_\omega} \, k_z  \right) \right\} - 1}{i \left( \frac{\omega}{c} - \sqrt{\varepsilon_\omega} \, k_z  \right)} \, e^{ -i \sqrt{\varepsilon_\omega} \, k_x x_0} 
\end{equation*}
\begin{equation}
    \label{I.plate.Durand}
    \times
    \frac{ \exp \left\{x_0 \sqrt{  \left(\frac{\omega}{v\gamma}\right)^2 + \left[\left( \frac{\omega}{v} - \sqrt{\varepsilon_\omega} \, k_z \right)\tan\psi - \sqrt{\varepsilon_\omega} \, k_y \right]^2} \right\} }
    {-ik_x + \sqrt{  \left(\frac{\omega}{v\gamma}\right)^2 + \left[\left( \frac{\omega}{v} - \sqrt{\varepsilon_\omega} \, k_z \right)\tan\psi - \sqrt{\varepsilon_\omega} \, k_y \right]^2}}
\end{equation}
\begin{equation*}
    \times
    \left\{ -\mathbf e_x +
    i \frac{\left( \frac{\omega}{v} - \sqrt{\varepsilon_\omega} \, k_z \right)\tan\psi - \sqrt{\varepsilon_\omega} \, k_y}
    { \sqrt{  \left(\frac{\omega}{v\gamma}\right)^2 +\left[\left( \frac{\omega}{v} - \sqrt{\varepsilon_\omega} \, k_z \right)\tan\psi - \sqrt{\varepsilon_\omega} \, k_y \right]^2}} \,
    \mathbf e_y \right\} \,.
\end{equation*}
Substitution of (\ref{I.plate.Durand}) into (\ref{spectral.angular.D}) leads to the result that coinsides with (13) in \cite{Tishch.2004}. The radiation cone, like in the case (\ref{I.eik.plate.0}), (\ref{delta.function.0}), (\ref{I.eik.plate.2}), is determined by the equality to zero of the value in the square brackets, which, however, is distinct from the same in (\ref{I.eik.plate.2}), and leads to the condition  
\begin{equation}\label{delta.function.5}
k_{y'} = \frac{1}{\sqrt{\varepsilon_\omega}} \frac{\omega}{v} \sin\psi 
\end{equation}
different from (\ref{delta.function.3}). The comparison of the spectral-angular density of DR computed using equations (\ref{I.eik.plate.2}) and (\ref{I.plate.Durand}) is presented in the figure \ref{our.vs.Tishch} (a) and (b) for the case $\varepsilon_\omega = 0.95$. We see that the cone's opening angle is very sensitive to the value $\varepsilon_\omega$. However, it is in controversy to Cherenkov interpretation of the conical effect discussed in the section \ref{interpretation}. 

This difficulty is fixed after account of the radiation refraction on the plate's boundary, as it has been made in \cite{Tishch.2}. To this goal we have to substitute (in our denominations)
\begin{equation*}
\sqrt{\varepsilon_\omega} \varkappa_x \to  \varkappa_x \,,
\end{equation*}
\begin{equation}\label{zamena}
\sqrt{\varepsilon_\omega} \, (\varkappa_z \sin\psi + \varkappa_y \cos\psi) \to  \varkappa_z \sin\psi + \varkappa_y \cos\psi  \quad \mbox{(that is} \ \  \sqrt{\varepsilon_\omega} \, \varkappa_{y'} \to  \varkappa_{y'} ) \,,
\end{equation}
\begin{equation*}
\sqrt{\varepsilon_\omega} \, (\varkappa_z \cos\psi -  \varkappa_y \sin\psi) \to  \sqrt{\varepsilon_\omega - 1 + (\varkappa_z \cos\psi -  \varkappa_y \sin\psi)^2} 
\end{equation*}
for the radiation that exits the target through the forward and backward faces. Under that, only the waves meeting the condition
\begin{equation}\label{poln.vnut}
(\varkappa_z \cos\psi - \varkappa_y \sin\psi)^2 > 1- \varepsilon_\omega
\end{equation}
can leave the target (condition (27) in \cite{Tishch.2}). The ranges permitted by (\ref{poln.vnut}) are located in figure \ref{geometry.1} on the left and right sides out of the green surfaces.

After the change (\ref{zamena}) the cone condition (\ref{delta.function.5}) transforms into (\ref{delta.function.3}), and the change in (\ref{I.plate.Durand}) with the following substitution into (\ref{spectral.angular.D}) leads to the result (30) in \cite{Tishch.2}. The last result in the range permitted by the condition (\ref{poln.vnut}) is very close to our (\ref{spectral.angular.1})--(\ref{spectral.angular.2}), see figure \ref{our.vs.Tishch} (d).

The range prohibited by the condition (\ref{poln.vnut}) could be filled with the radiadiation refracted by the upper face of the plate. The description of such radiation could be found in \cite{Tishch.1}, however, the authors outline that their results are valid for a thin wide plate (the width $a$ is much larger than the plate's thickness along the $x$ axis). 

\section{Conclusion}
\label{sec:conclusion}

The unified description of diffraction and transition radiation arising in the X-ray domain under oblique incidence of relativistic particle on the semi-infinite plate parallel to its upper face is presented. The calculations are based on the simple variant of eikonal approximation developed earlier \cite{arsa.2006, pov.2007}. This approach accounts the evolution of the electric field created in the target by the propagating particle. 

The comparison with the other authors' results \cite{Tishch.2004, Tishch.2} demonstrates that our approach automatically takes into account the radiation refraction on the target boundary, and also describes the radiation leaving the target through the forward and through the upper faces in a unified way.



\end{document}